\documentclass[useAMS,usenatbib,psfig]{mn2e}

\usepackage{amssymb}
\usepackage{times}
\usepackage{color}
\usepackage{ulem}
\usepackage{graphicx}

\title[The power of the jets] {The power of the jets accelerated by the coronal magnetic field}
\author[X. Cao]
{Xinwu Cao$^{1,2,3}$\\
$^1$ Shanghai Astronomical Observatory, Chinese Academy of Sciences,
80 Nandan Road, Shanghai, 200030, China; E-mail: cxw@shao.ac.cn\\
$^2$ Key Laboratory of Radio Astronomy, Chinese Academy of Sciences,
210008 Nanjing, China\\
$^3$ University of Chinese Academy of Sciences, Beijing 100049,
China\\}

\date{Accepted 2017 October 11 . Received 2017 October 11 ; in original form 2017 January 01}

\pagerange{\pageref{firstpage}--\pageref{lastpage}} \pubyear{}

\begin{document}

\maketitle \label{firstpage}

\begin{abstract}
It was suggested that the large scale magnetic field can be dragged
inwards efficiently by the corona above the disc, i.e., the so called
``coronal mechanism" (Beckwith, Hawley, \& Krolik 2009), which
provides a way to solve the difficulty of field advection in a
geometrically thin accretion disc. In this case, the magnetic
pressure should be lower than the gas pressure in the corona. We
estimate the maximal power of the jets accelerated by the magnetic
field advected by the corona. The Blandford-Payne (BP) jet power is
found always to be higher than the Blandford-Znajek (BZ) jet power,
except for a rapidly spinning black hole with $a\ga 0.8$. The
maximal jet power is always low, less than 0.05 Eddington
luminosity, even for an extreme Kerr black hole, which is
insufficient for the observed strong jets in some blazars with jet
power $\sim 0.1-1$ Eddington luminosity (or even higher). It implies that these powerful jets cannot be accelerated by
the coronal field. We suggest
that, the magnetic field dragged inward by the accretion disc with
magnetically outflows may accelerate the jets (at least for the most
powerful jets, if not all) in the blazars.
\end{abstract}


\section{Introduction}\label{intro}

Jets are believed to be accelerated by the large scale magnetic
field through either the Blandford-Znajek (BZ) or the Blandford
Payne (BP) mechanisms
\citep*[][]{1977MNRAS.179..433B,1982MNRAS.199..883B}. The kinetic
power of a spinning black hole or the gas in the accretion disc is
tapped into the jets with the co-rotating large scale magnetic
field. The numerical simulations show that the net field flux is
necessary for jet formation \citep*[][]{2016MNRAS.460.3488S}, and
the large scale magnetic field accelerating jets may probably be
formed by the advection of the external weak field (e.g., the field
threading the interstellar medium)
\citep*[][]{1974Ap&SS..28...45B,1976Ap&SS..42..401B,1989ASSL..156...99V,1994MNRAS.267..235L}.

Relativistic jets have been observed in radio-loud quasars. The
power of the jets can be as high as the Eddington luminosity in some
radio-loud quasars
\citep*[e.g.,][]{2009MNRAS.396..984G,2014Natur.515..376G}. Quasars
are accreting at high rates, which may probably contain standard
thin accretion discs, which implies that the weak external magnetic
field should be dragged inwards by the thin accretion disc in order
to form a sufficient strong field to accelerate relativistic jets
near the black hole. In a steady case, the advection of the field is
balanced with the magnetic diffusion in the accretion disc. The
radial velocity of a conventional turbulent accretion disc is mainly
regulated by the kinematic viscosity $\nu$, and therefore the field
advection in the disc is sensitive to the magnetic Prandtl number
$P_{\rm m}=\eta/\nu$ ($\eta$ is the magnetic diffusivity). Both the
simple estimate of the order of magnitude \citep*[][]{parker1979}
and the numerical simulations show that the magnetic Prandtl number
is always around unity
\citep*[e.g.,][]{2003A&A...411..321Y,2009A&A...504..309L,2009A&A...507...19F,2009ApJ...697.1901G}.
It was found that the advection of the external field is rather
inefficient in the geometrically thin accretion disc ($H/R\ll 1$)
\citep*[][]{1994MNRAS.267..235L}, because of its small radial
velocity. This means that the field in the inner region of the disc
is not much stronger than the external weak field
\citep*[][]{1994MNRAS.267..235L}, which is unable to accelerate
strong jets in radio-loud quasars.

A few mechanisms were suggested to alleviate the difficulty of field
advection in the thin discs
\citep*[][]{2005ApJ...629..960S,2009ApJ...701..885L,2012MNRAS.424.2097G,2013MNRAS.430..822G,2013ApJ...765..149C}.
It was suggested that the external field can be dragged efficiently
inwards by the hot corona above the disc, i.e., the so called
``coronal mechanism" \citep*[see][for the
details]{2009ApJ...707..428B}. The radial velocity of the gas above
the disc can be larger than that at the midplane of the disc, which
partially solve the problem of inefficient field advection in the
thin disc
\citep*[][]{2009ApJ...701..885L,2012MNRAS.424.2097G,2013MNRAS.430..822G}.
Alternatively, \citet{2013ApJ...765..149C} suggested that the radial
velocity of the disc is significantly increased, if the most angular
momentum of the gas in the thin disc is removed by the magnetically
driven outflows, and therefore the external field can be
significantly enhanced in the inner region of the thin disc with
magnetic outflows.

In this work, we estimate the maximal strength of the field dragged
inwards by the hot corona, and then derive the maximal power of the
jets accelerated either by the BP or BZ mechanisms. In Section
\ref{b_of_corona}, we estimate the strength of the large scale
magnetic field dragged inwards by the corona. The maximal power of
the jets accelerated by this magnetic field is derived in Section
\ref{jet_power}. The last section contains the results and
discussion.

\section{Magnetic field dragged inwards by the hot corona}\label{b_of_corona}

The detailed properties of the corona above the the disc is still
unclear, though it has been extensively studied by many authors
\citep*[e.g.,][]{1979ApJ...229..318G,1991ApJ...380L..51H,1993ApJ...413..507H,1994ApJ...436..599S,2001ApJ...546..966K,
2009MNRAS.394..207C}. However, it is quite
certain that the hot corona is geometrically thick and optically
thin. In this paper, we use relative thickness $\tilde{H}_{\rm
c}=H_{\rm c}/R$ and the optical depth $\tau_{\rm c}$ to describe the
corona above the accretion disc. The optical depth of the corona in
the vertical direction is
\begin{equation}
\tau_{\rm c}=\rho_{\rm c}H_{\rm c}\kappa_{\rm T}, \label{tau_c}
\end{equation}
where $\rho_{\rm c}$ is the density of the corona, $H_{\rm c}$ is
the corona thickness, and $\kappa_{\rm T}=0.4~{\rm g}^{-1}{\rm
cm}^2$ is the Compton scattering opacity.

{The gas pressure of the corona is
%\begin{equation}
%p_{\rm c}=\rho_{\rm c}c_{\rm s,c}^2=\rho_{\rm c}(H_{\rm
%c}\Omega_{\rm K})^2=\tau_{\rm c}H_{\rm c}\Omega_{\rm K}^2\kappa_{\rm
%T}^{-1},\label{p_c}
%\end{equation}
\begin{equation}
p_{\rm c}={\frac {\rho_{\rm c}}{2}}\left({\frac {H_{\rm c}}{R}}\right)^2{\frac {L_*^2}{R^2}},
\label{p_c}
\end{equation}
where $a$ is the spin parameter of the black hole, and
\begin{equation}
L_{*}^2=L^2-a^2(E^2-1),\label{l_star}
\end{equation}
\citep*[see][for the details]{1997ApJ...479..179A}.
The conserved angular momentum of the gas $L=u_\phi$, and the conserved energy $E=-u_t$.}

%where the sound speed $c_{\rm s,c}=H_{\rm c}\Omega_{\rm K}$ and%
%Equation (\ref{tau_c}) are used.

We use a parameter $\beta$ to describe the magnetic field strength
in the corona,
\begin{equation}
p_{\rm m}={\frac {B_{z}^2}{8\pi}}=\beta p_{\rm c},\label{p_m}
\end{equation}
where $B_z$ is the strength of the vertical component of the field,
and $\beta<1$ is required for the field advected by the corona. {The
field strength of the corona is
\begin{equation} B_z=4.37\times
10^8\beta^{1/2}\tau_{\rm c}^{1/2}\tilde{H}_{\rm
c}^{1/2}m^{-1/2}r^{-3/2}L_{*}^2~{\rm Gauss}, \label{b_z}
\end{equation}
where
\begin{equation}
\tilde{H}_{\rm c}={\frac {H_{\rm c}}{R}},~~ m={\frac {M_{\rm
bh}}{M_\odot}},~~{\rm and}~~ r={\frac {Rc^2}{GM_{\rm bh}}}.
\end{equation}}

\section{Jet power}\label{jet_power}

The power of the jets driven by the field threading a spinning black
hole is \citep*[][]{1982MNRAS.198..345M,1997MNRAS.292..887G}
\begin{equation}
P_{\rm BZ}={\frac {1}{32}}\omega_{\rm F}^2B_{\rm h}^2R_{\rm
h}^2ca^2, \label{p_bz}
\end{equation}
where $B_{\rm h}$ is the field strength at the black hole horizon
$R_{\rm h}$, and $\omega_{\rm F}^2$ describes the effects of the
angular velocity $\Omega_{\rm F}$ of the field lines relative to
black hole angular velocity. The BZ jet power is maximized if
$\omega_{\rm F}=1/2$ is adopted
\citep*[][]{1982MNRAS.198..345M,1997MNRAS.292..887G}. Substituting
Equation (\ref{b_z}) into Equation (\ref{p_bz}), {we have
\begin{equation} P_{\rm BZ}=3.9\times 10^{36}\omega_{\rm
F}^2\beta\tau_{\rm c}mr_{\rm h}^{-1}\tilde{H}_{\rm c}L_{*}^2(r_{\rm
h})a^2~{\rm erg}, \label{p_bz2}
\end{equation}
or
\begin{equation}
{\frac {P_{\rm BZ}}{L_{\rm Edd}}}={\frac {P_{\rm BZ}}{1.251\times
10^{38}m}}=3.12\times 10^{-2}\omega_{\rm
F}^2\beta\tau_{\rm c}r_{\rm h}^{-1}\tilde{H}_{\rm c}L_{*}^2(r_{\rm h})a^2. \label{p_bz3}
\end{equation}}
{The gas falls almost freely onto the black hole, and the angular
momentum of the gas at the black hole horizon $L(r_{\rm h})\la
L(r_{\rm ms})$ ($r_{\rm ms}$ is the radius of the marginal stable
circular orbits). The conserved energy $|E|$ does not deviate much
from the unity, so the second term in Equation (\ref{l_star}) is
always negligible. The global solution of a relativistic accretion
flow surrounding a Kerr black hole shows that the angular momentum
of the gas at the black hole horizon is slightly lower than that at
the radius of the marginal stable circular orbits
\citep*[e.g.,][]{1996ApJ...471..762A,1998ApJ...498..313G,2000ApJ...534..734M}.
As we intend to estimate the maximal jet power, we conservatively
adopt $L_{*}=L_{\rm K}(r_{\rm ms})$ in the estimate of the strength
of the field at the black hole horizon.}

The power of the jets launched by the BP mechanism is estimated as
\citep*[][]{1999ApJ...512..100L}
\begin{equation}
P_{\rm BP}\sim {\frac {B_zB_\phi^{\rm s}}{2\pi}}R_{\rm j}\Omega\pi
R_{\rm j}^2, \label{p_bp}
\end{equation}
where $R_{\rm j}$ is the typical radius of the jet formation region
in the corona, $\Omega$ is the angular velocity of the gas in the
corona, $B_\phi^{\rm s}$ is the azimuthal component of the field at
the corona surface, and $B_\phi^{\rm s}=\xi_\phi B_z$. The ratio
$\xi_\phi\la 1$ is required \citep*[see][for the detailed
discussion]{1999ApJ...512..100L}. Substituting Equation (\ref{b_z})
into Equation (\ref{p_bp}), we derive the BP power of the jets as
\begin{equation}
P_{\rm BP}\sim 3.13\times 10^{37}\xi_\phi\tilde{\Omega}r_{\rm
j}^{-1/2}m\beta\tau_{\rm c}\tilde{H}_{\rm c}~{\rm erg~s}^{-1},
\label{p_bp2}
\end{equation}
or
\begin{equation}
{\frac {P_{\rm BP}}{L_{\rm Edd}}}={\frac {P_{\rm
BP}}{1.251\times10^{38}m}}\sim 0.25 \xi_\phi\tilde{\Omega}r_{\rm
j}^{-1/2}\beta\tau_{\rm c}\tilde{H}_{\rm c}. \label{p_bp3}
\end{equation}
As the most gravitational power is released in the inner region of
the accretion disc within the radius of $\sim 2R_{\rm ms}$
\citep*[][]{1973A&A....24..337S}, we adopt $R_{\rm j}=2R_{\rm ms}$
in all the estimates of the BP power $P_{\rm BP}$.

\section{Results and discussion}\label{discussion}

%We plot the ratios $r_{\rm ms}/r_{\rm h}$ as functions of the black
%hole spin parameter $a$ in Figure \ref{rat_rm_rh}. The ratio
%decreases with increasing the black hole spin parameter $a$ if the
%corona rotates with the black hole in prograde orbits, while the
%ratio increases with $a$ if the corona is in the retrograde orbits
%with the spinning black hole.

\begin{figure}
  \includegraphics[width=0.48\textwidth]{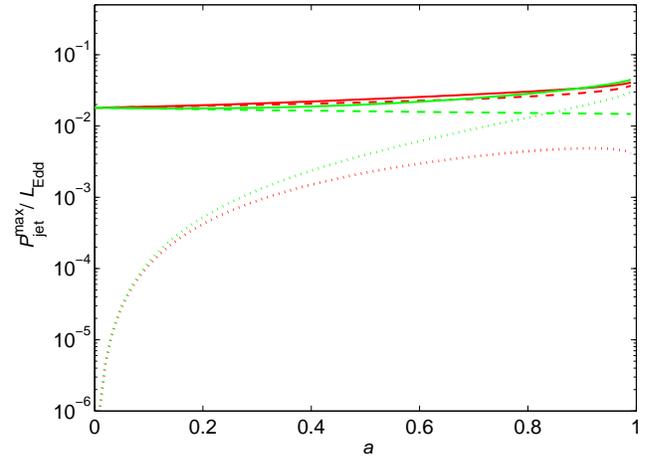}
\caption{The maximal jet power as functions of the black spin
parameter $a$. The red lines represents the results for the prograde
orbits, while the green lines are for the retrograde orbits. The
dotted lines are the maximal BZ jet power, while the dashed lines
are the maximal BP power. {The solid lines are the total jet power,
$P_{\rm jet}=P_{\rm BZ}+P_{\rm BP}$.} } \label{p_bz_bp}
\end{figure}

In the case of the magnetic field dragged inwards by the corona, the
magnetic pressure is required to be lower than the gas pressure in
the corona, i.e., $\beta<1$. In order to estimate the maximal power
of the jets driven by the BP or BZ mechanisms, we adopt $\omega_{\rm
F}=1/2$ \citep*[][]{1997MNRAS.292..887G}, $\beta=1$, $\xi_\phi=1$,
$\xi_{\rm pl}=(r_{\rm ms}/r_{\rm h})^2$, and $\tilde{\Omega}=1$,
while the typical values of the corona parameters, $\tau_{\rm
c}=0.5$, and $\tilde{H}_{\rm c}=0.5$ are adopted in the estimates
\citep*[e.g.,][]{2009MNRAS.394..207C}. The maximal jet powers as
functions of the black spin parameter $a$ are plotted in Figure
\ref{p_bz_bp}.

{The maximal BZ power for the corona with prograde orbits
surrounding the spinning black hole is significantly less than 0.01
Eddington luminosity, which is always lower than the BP power for
any value of $a$. This is similar to the accretion disc case
\citep*[see the discussion in][]{1999ApJ...512..100L}. The situation
is quite different in the retrograde orbit case. For the corona with
retrograde orbits, the maximal BP power is higher than the BZ power
except for a rapidly spinning black hole ($a\ga0.8$) (see Figure
\ref{p_bz_bp}). {The maximal BZ power for the retrograde orbit case
is higher than that for the prograde case, which is qualitatively
consistent with recent numerical simulations, though based on
different assumptions \citep*[][]{2015MNRAS.446L..61P}.} The maximal
jet power increases with the spin parameter $a$, which is found to
be less than $\sim 0.05~L_{\rm Edd}$ even for an extreme Kerr black
hole.} However, it is still insufficient for the observed strong
jets with power around the Eddington luminosity in some blazars
\citep*[e.g.,][]{2009MNRAS.396..984G,2014Natur.515..376G,2014ApJS..215....5K,2015ApJ...807...51Z}.
It implies that these powerful jets cannot be accelerated by the
coronal field. We suggest that, the magnetic field dragged inwards by
the accretion disc with magnetically outflows may accelerate the
jets (at least for the most powerful jets, if not all) in the
blazars
\citep*[][]{2013ApJ...765..149C,2014ApJ...788...71L,2016ApJ...833...30C}.
In this case, hot gas/corona above the thin disc may help launching
outflows/jets \citep*[][]{2013ApJ...770...31W,2014ApJ...783...51C},
though the field is not formed through ``coronal mechanism". For
those less powerful jets in radio galaxies, there is evidence that
the jets may be closely related to the ADAFs surrounding spinning
black holes
\citep*[e.g.,][]{2013MNRAS.436.1278W,2017MNRAS.470..612F}.

The accretion disc may be vertically compressed by the advected
large scale magnetic field
\citep*[][]{2002A&A...385..289C,2011ApJ...737...94C}. This may also
be the case for the corona case discussed in this paper. If this
effect is properly considered, the maximal magnetic field strength,
and then the maximal jet power (either the BZ or the BP power), will
be lower than the values derived in this paper, when the relative
corona thickness $\tilde{H}_{\rm c}$ is decreased (see Equation
\ref{b_z}).

In the strong magnetic field case, the accretion disc may be
magnetically arrested by the advected field
\citep*[][]{2003PASJ...55L..69N,2008ApJ...677..317I,2011ApJ...737...94C,2011MNRAS.418L..79T,2014ApJ...789..129C}.
{The numerical simulations were carried out for the geometrically
thick accretion flows, i.e., the advection dominated accretion flows
(ADAFs)
\citep{2008ApJ...677..317I,2011MNRAS.418L..79T,2012MNRAS.423L..55T},
which show that the jet efficiency $\eta_{\rm jet}$ can be higher
than 100\% under the certain circumstances (the jet efficiency
$\eta_{\rm jet}=P_{\rm jet}/\dot{M}c^2$, and $\dot{M}$ is the mass
accretion rate of the accretion flow). The ADAF is suppressed when
the dimensionless mass accretion rate $\dot{m}$ is greater than a
critical value $\dot{m}_{\rm crit}$ ($\dot{m}=\dot{M}/\dot{M}_{\rm
Edd}$), which is suggested to be around $0.01$
\citep*[][]{1995ApJ...452..710N}. This implies that the maximal jet
power for an ADAF should be $\sim 0.1~L_{\rm Edd}$. }

{The advection of the external large scale magnetic field is
efficient in the ADAF due to its large radial velocity, and a very
strong magnetic field is formed near the black to arrest the
accretion flow
\citep*[][]{2003PASJ...55L..69N,2008ApJ...677..317I,2011ApJ...737...94C,2011MNRAS.418L..79T,2014ApJ...789..129C}.
This is different from the accretion disc-corona case considered in
this paper. In the accretion disc-corona system, the external large
scale field is dragged inward by the hot corona above/below the
disc. Similar to an ADAF, the radial velocity of the hot corona is
much higher than that of the thin disc, and the field is advected
inward efficiently. However, the dragged field lines will have to
pass through the geometrically thin disc located between the
coronae. The field dragged by the corona may be diffused in the thin
disc, and therefore it is sceptical whether the field can be
developed to be so strong to arrest the accretion disc-corona. In
this case, the maximal strength of the field advected by the corona
is still limited by the condition $\beta\la 1$. }

%Fig. 1

\section*{Acknowledgments}
I thank the referee for his/her helpful comments/suggestions. This
work is supported by the NSFC (grants 11233006 and 11773050), the
CAS grant (QYZDJ-SSW-SYS023), and Shanghai Municipality.


\begin{thebibliography}{}

\bibitem[\protect\citeauthoryear{Abramowicz et al.}{1996}]{1996ApJ...471..762A} Abramowicz M.~A., Chen X.-M., Granath M., Lasota J.-P., 1996, ApJ, 471, 762

\bibitem[\protect\citeauthoryear{Abramowicz, Lanza, \& Percival}{1997}]{1997ApJ...479..179A} Abramowicz M.~A., Lanza A., Percival M.~J., 1997, ApJ, 479, 179

\bibitem[\protect\citeauthoryear{Beckwith, Hawley, \& Krolik}{2009}]{2009ApJ...707..428B} Beckwith K., Hawley J.~F., Krolik J.~H., 2009, ApJ, 707, 428

\bibitem[\protect\citeauthoryear{Bisnovatyi-Kogan \& Ruzmaikin}{1974}]{1974Ap&SS..28...45B} Bisnovatyi-Kogan G.~S., Ruzmaikin A.~A., 1974, Ap\&SS, 28, 45

\bibitem[\protect\citeauthoryear{Bisnovatyi-Kogan \& Ruzmaikin}{1976}]{1976Ap&SS..42..401B} Bisnovatyi-Kogan G.~S., Ruzmaikin A.~A., 1976, Ap\&SS, 42, 401

\bibitem[\protect\citeauthoryear{Blandford \& Payne}{1982}]{1982MNRAS.199..883B} Blandford R.~D., Payne D.~G., 1982, MNRAS, 199, 883

\bibitem[\protect\citeauthoryear{Blandford \& Znajek}{1977}]{1977MNRAS.179..433B} Blandford R.~D., Znajek R.~L., 1977, MNRAS, 179, 433

\bibitem[\protect\citeauthoryear{Cao}{2009}]{2009MNRAS.394..207C} Cao X., 2009, MNRAS, 394, 207

\bibitem[\protect\citeauthoryear{Cao}{2011}]{2011ApJ...737...94C} Cao X., 2011, ApJ, 737, 94

\bibitem[\protect\citeauthoryear{Cao}{2014}]{2014ApJ...783...51C} Cao X., 2014, ApJ, 783, 51

\bibitem[\protect\citeauthoryear{Cao}{2016}]{2016ApJ...833...30C} Cao X., 2016, ApJ, 833, 30

\bibitem[\protect\citeauthoryear{Cao, Liang, \& Yuan}{2014}]{2014ApJ...789..129C} Cao X., Liang E.-W., Yuan Y.-F., 2014, ApJ, 789, 129

\bibitem[\protect\citeauthoryear{Cao \& Spruit}{2002}]{2002A&A...385..289C} Cao X., Spruit H.~C., 2002, A\&A, 385, 289

\bibitem[\protect\citeauthoryear{Cao \& Spruit}{2013}]{2013ApJ...765..149C} Cao X., Spruit H.~C., 2013, ApJ, 765, 149

\bibitem[\protect\citeauthoryear{Feng \& Wu}{2017}]{2017MNRAS.470..612F} Feng J., Wu Q., 2017, MNRAS, 470, 612

\bibitem[\protect\citeauthoryear{Fromang \& Stone}{2009}]{2009A&A...507...19F} Fromang S., Stone J.~M., 2009, A\&A, 507, 19

\bibitem[\protect\citeauthoryear{Galeev, Rosner, \& Vaiana}{1979}]{1979ApJ...229..318G} Galeev A.~A., Rosner R., Vaiana G.~S., 1979, ApJ, 229, 318

\bibitem[\protect\citeauthoryear{Gammie \& Popham}{1998}]{1998ApJ...498..313G} Gammie C.~F., Popham R., 1998, ApJ, 498, 313

%\bibitem[\protect\citeauthoryear{Garofalo}{2009}]{2009ApJ...699..400G} Garofalo D., 2009, ApJ, 699, 400

\bibitem[\protect\citeauthoryear{Ghisellini et al.}{2014}]{2014Natur.515..376G} Ghisellini G., Tavecchio F., Maraschi L., Celotti A., Sbarrato T., 2014, Natur, 515, 376

\bibitem[\protect\citeauthoryear{Ghosh \& Abramowicz}{1997}]{1997MNRAS.292..887G} Ghosh P., Abramowicz M.~A., 1997, MNRAS, 292, 887

\bibitem[\protect\citeauthoryear{Gu, Cao, \& Jiang}{2009}]{2009MNRAS.396..984G} Gu M., Cao X., Jiang D.~R., 2009, MNRAS, 396, 984

\bibitem[\protect\citeauthoryear{Guan \& Gammie}{2009}]{2009ApJ...697.1901G} Guan X., Gammie C.~F., 2009, ApJ, 697, 1901

\bibitem[\protect\citeauthoryear{Guilet \& Ogilvie}{2012}]{2012MNRAS.424.2097G} Guilet J., Ogilvie G.~I., 2012, MNRAS, 424, 2097

\bibitem[\protect\citeauthoryear{Guilet \& Ogilvie}{2013}]{2013MNRAS.430..822G} Guilet J., Ogilvie G.~I., 2013, MNRAS, 430, 822

\bibitem[\protect\citeauthoryear{Lubow, Papaloizou, \& Pringle}{1994}]{1994MNRAS.267..235L} Lubow S.~H., Papaloizou J.~C.~B., Pringle J.~E., 1994, MNRAS, 267, 235

\bibitem[\protect\citeauthoryear{MacDonald \& Thorne}{1982}]{1982MNRAS.198..345M} MacDonald D., Thorne K.~S., 1982, MNRAS, 198, 345

\bibitem[\protect\citeauthoryear{Manmoto}{2000}]{2000ApJ...534..734M} Manmoto T., 2000, ApJ, 534, 734

\bibitem[\protect\citeauthoryear{Haardt \& Maraschi}{1991}]{1991ApJ...380L..51H} Haardt F., Maraschi L., 1991, ApJ, 380, L51

\bibitem[\protect\citeauthoryear{Haardt \& Maraschi}{1993}]{1993ApJ...413..507H} Haardt F., Maraschi L., 1993, ApJ, 413, 507

\bibitem[\protect\citeauthoryear{Igumenshchev}{2008}]{2008ApJ...677..317I} Igumenshchev I.~V., 2008, ApJ, 677, 317-326

\bibitem[\protect\citeauthoryear{Kang, Chen, \& Wu}{2014}]{2014ApJS..215....5K} Kang S.-J., Chen L., Wu Q., 2014, ApJS, 215, 5

\bibitem[\protect\citeauthoryear{Kawaguchi, Shimura, \& Mineshige}{2001}]{2001ApJ...546..966K} Kawaguchi T., Shimura T., Mineshige S., 2001, ApJ, 546, 966

\bibitem[\protect\citeauthoryear{Lesur \& Longaretti}{2009}]{2009A&A...504..309L} Lesur G., Longaretti P.-Y., 2009, A\&A, 504, 309

\bibitem[\protect\citeauthoryear{Li}{2014}]{2014ApJ...788...71L} Li S.-L., 2014, ApJ, 788, 71

%\bibitem[\protect\citeauthoryear{Liu, Mineshige, \& Ohsuga}{2003}]{2003ApJ...587..571L} Liu B.~F., Mineshige S., Ohsuga K., 2003, %ApJ, 587, 571

\bibitem[\protect\citeauthoryear{Livio, Ogilvie, \& Pringle}{1999}]{1999ApJ...512..100L} Livio M., Ogilvie G.~I., Pringle J.~E., 1999, ApJ, 512, 100

\bibitem[\protect\citeauthoryear{Lovelace, Rothstein, \& Bisnovatyi-Kogan}{2009}]{2009ApJ...701..885L} Lovelace R.~V.~E., Rothstein D.~M., Bisnovatyi-Kogan G.~S., 2009, ApJ, 701, 885

\bibitem[\protect\citeauthoryear{Manmoto}{2000}]{2000ApJ...534..734M} Manmoto T., 2000, ApJ, 534, 734

\bibitem[\protect\citeauthoryear{Narayan, Igumenshchev, \& Abramowicz}{2003}]{2003PASJ...55L..69N} Narayan R., Igumenshchev I.~V., Abramowicz M.~A., 2003, PASJ, 55, L69

\bibitem[\protect\citeauthoryear{Narayan \& Yi}{1995}]{1995ApJ...452..710N} Narayan R., Yi I., 1995, ApJ, 452, 710

\bibitem[\protect\citeauthoryear{Parfrey, Giannios, \&
Beloborodov}{2015}]{2015MNRAS.446L..61P} Parfrey K., Giannios D.,
Beloborodov A.~M., 2015, MNRAS, 446, L61


\bibitem[\protect\citeauthoryear{Parker}{1979}]{parker1979} Parker, E.~N.\ 1979, in Chapter 17, Cosmical Magnetic
Fields (Oxford:Clarendon Press)

%\bibitem[\protect\citeauthoryear{Reynolds, Garofalo, \& Begelman}{2006}]{2006ApJ...651.1023R} Reynolds C.~S., Garofalo D., Begelman M.~C., 2006, ApJ, 651, 1023

\bibitem[\protect\citeauthoryear{Salvesen et al.}{2016}]{2016MNRAS.460.3488S} Salvesen G., Armitage P.~J., Simon J.~B., Begelman M.~C., 2016, MNRAS, 460, 3488

\bibitem[\protect\citeauthoryear{Shakura \& Sunyaev}{1973}]{1973A&A....24..337S} Shakura N.~I., Sunyaev R.~A., 1973, A\&A, 24, 337

\bibitem[\protect\citeauthoryear{Spruit \& Uzdensky}{2005}]{2005ApJ...629..960S} Spruit H.~C., Uzdensky D.~A., 2005, ApJ, 629, 960

\bibitem[\protect\citeauthoryear{Svensson \& Zdziarski}{1994}]{1994ApJ...436..599S} Svensson R., Zdziarski A.~A., 1994, ApJ, 436, 599

\bibitem[\protect\citeauthoryear{Tchekhovskoy \& McKinney}{2012}]{2012MNRAS.423L..55T} Tchekhovskoy A., McKinney J.~C., 2012, MNRAS, 423, L55

\bibitem[\protect\citeauthoryear{Tchekhovskoy, Narayan, \& McKinney}{2011}]{2011MNRAS.418L..79T} Tchekhovskoy A., Narayan R., McKinney J.~C., 2011, MNRAS, 418, L79

\bibitem[\protect\citeauthoryear{van Ballegooijen}{1989}]{1989ASSL..156...99V} van Ballegooijen, A.~A. 1989, in Accretion
Disks and Magnetic Fields in Astrophysics, ed. G. Belvedere (ASSL
Vol. 156; Dordrecht: Kluwer), 99

\bibitem[\protect\citeauthoryear{Wu et al.}{2013}]{2013ApJ...770...31W} Wu Q., Cao X., Ho L.~C., Wang D.-X., 2013, ApJ, 770, 31

\bibitem[\protect\citeauthoryear{Wu, Yan, \& Yi}{2013}]{2013MNRAS.436.1278W} Wu Q., Yan H., Yi Z., 2013, MNRAS, 436, 1278

\bibitem[\protect\citeauthoryear{Yousef, Brandenburg, \& R{\"u}diger}{2003}]{2003A&A...411..321Y} Yousef T.~A., Brandenburg A., R{\"u}diger G., 2003, A\&A, 411, 321

\bibitem[\protect\citeauthoryear{Zhang et al.}{2015}]{2015ApJ...807...51Z} Zhang J., Xue Z.-W., He J.-J., Liang E.-W., Zhang S.-N., 2015, ApJ, 807, 51

\end{thebibliography}
\end{document}